\UseRawInputEncoding
\documentclass[twocolumn,showpacs,aps,prl,superscriptaddress,floatfix]{revtex4}

\usepackage{graphicx}
\usepackage{verbatim}
\usepackage{dcolumn}
\usepackage{amsmath}
\usepackage{epsfig}

\newcommand{\BaBarNumber}    {21/003}
\newcommand{\BaBarType}      {PUB}  
\newcommand{\SLACPubNumber}  {17617}

\input babarsym

\def\emu {\ensuremath{e\mu}\xspace}

\begin{document}

\pagestyle{plain}

\begin{flushleft}
\babar-\BaBarType-\BaBarNumber \\ SLAC-PUB-\SLACPubNumber  \\
\end{flushleft}



\title{{\large \bf Search for Lepton Flavor Violation in $\Upsilon (3S)\rightarrow e^{\pm}\mu^{\mp}$}}

\author{J.~P.~Lees}
\author{V.~Poireau}
\author{V.~Tisserand}
\affiliation{Laboratoire d'Annecy-le-Vieux de Physique des Particules (LAPP), Universit\'e de Savoie, CNRS/IN2P3,  F-74941 Annecy-Le-Vieux, France}
\author{E.~Grauges}
\affiliation{Universitat de Barcelona, Facultat de Fisica, Departament ECM, E-08028 Barcelona, Spain }
\author{A.~Palano}
\affiliation{INFN Sezione di Bari, I-70126 Bari, Italy}
\author{G.~Eigen}
\affiliation{University of Bergen, Institute of Physics, N-5007 Bergen, Norway }
\author{D.~N.~Brown}
\author{Yu.~G.~Kolomensky}
\affiliation{Lawrence Berkeley National Laboratory and University of California, Berkeley, California 94720, USA }
\author{M.~Fritsch}
\author{H.~Koch}
\author{T.~Schroeder}
\affiliation{Ruhr Universit\"at Bochum, Institut f\"ur Experimentalphysik 1, D-44780 Bochum, Germany }
% ...
\collaboration{The \babar\ Collaboration}
\noaffiliation



\begin{abstract}


We report on the first search for electron-muon lepton flavor violation (LFV) in the decay of
 a $b$~quark and $b$~antiquark bound state.
We look for the LFV decay $\Upsilon (3S)\rightarrow e^{\pm}\mu^{\mp}$
in a sample of  118~million $\Upsilon (3S)$ mesons from 
27~fb$^{-1}$ of data collected with the \babar\ detector at the SLAC PEP-II $e^+e^-$ collider 
operating with a 10.36~GeV center-of-mass energy. No evidence for a signal is found and
we set a limit on the branching fraction
$\mathcal{B}(\Upsilon(3S)\rightarrow e^{\pm}\mu^{\mp})<3.6\times10^{-7}\mathrm{at~ 90\%~ CL}$.
This result can be interpreted as a limit $\Lambda_{NP}/g^2_{NP} > 80~$TeV
 on the energy scale $\Lambda_{NP}$ divided by the coupling-squared $g^2_{NP}$
 of relevant new physics.


\end{abstract}

\maketitle

In the standard model (SM), the three lepton flavors (electron, muon, tau)
are carried by the charged leptons ($e^-$, $\mu^-$, and $\tau^-$) and their associated neutrinos
 ($\nu_e$, $\nu_{\mu}$, $\nu_{\tau}$). Were it not for the fact that
 neutrinos oscillate from one flavor to another, lepton flavor would be strictly
  conserved in all reactions in the SM. Although  mixing between the neutrino flavor eigenstates
 permits  charged lepton flavor violating (LFV) processes at higher-order, these are extremely suppressed in the SM
by powers of the small neutrino masses. Therefore, observation of charged LFV  
would be a clear signature of new physics (NP),
and placing experimentally stringent limits on the branching fractions of such processes 
tightly constrains NP models.
Searches for electron-tau and muon-tau LFV 
 in  decays of bound states of a $b$~quark and  $b$~antiquark ($b{\bar b}$)
have yielded no evidence of a signal and upper limits ranging from 3.1$\times 10^{-6}$ to 6.0$\times 10^{-6}$ 
 on their branching fractions have been set~\cite{Zyla:2020zbs}.
 This paper describes the first search for electron-muon  LFV 
 in the decay of a $b{\bar b}$ bound-state.\\
\indent Indirect theoretical constraints on LFV decays of vector (i.e., $\hbox{spin}=1, \hbox{parity}=-1$) $b{\bar b}$ bound states
 (referred to as the $\Upsilon(nS)$ mesons, $n=1,2,3,4...$) can be derived using an argument based 
on the non-observation of LFV decays of the muon in conjunction with unitarity considerations~\cite{ref:nussinov}.
In these calculations, it is assumed that a virtual $\Upsilon$ meson 
  could potentially contribute to the muon LFV decay. 
The most stringent indirect bound on electron-muon LFV decays of the $\Upsilon (3S)$ (with mass 
$M_{\Upsilon(3S)}=10.36$~GeV) obtained in this way
is $\mathcal{B}(\Upsilon(3S) \rightarrow e^{\pm}\mu^{\mp}) \le 2.5 \times 10^{-8}$, which uses the 
reported limit on the branching fraction $\mathcal{B}(\mu$ $\rightarrow 3e) < 1.0\times 10^{-12}$~\cite{ref:bellgardt}.
Using LFV limits from $\mu$-$e$ conversions, Ref.~\cite{ref:Gutsche} sets an upper bound at $3.9\times 10^{-6}$.
However, it has been noted in Ref.~\cite{ref:nussinov} that
the size of the vector boson exchange contribution to the $\mu \rightarrow 3e$ decay 
amplitude can be significantly reduced if there are kinematical suppressions. Such 
suppressions are possible when the effective vector boson couplings involve derivatives 
(or momentum factors). 
 This possibility means 
there could be effective tensor and pseudo-tensor LFV couplings in the $\mu \rightarrow 
3e$ decay, which would reduce the contribution of virtual $\Upsilon$(nS) bosons as they
only have vector couplings. Reference~\cite{ref:nussinov} estimates that the contribution 
of the virtual $\Upsilon(3S) \rightarrow e^{\pm} \mu^{\mp}$ to the $\mu \rightarrow 3e$ 
rate would be reduced by approximately 
${M_{\mu}^{2}}/{ (2 M_{\Upsilon(3S)}^{2})}$, 
leading to a re-calculated bound on $\mathcal{B}(\Upsilon (3S)\rightarrow e^{\pm}\mu^{\mp}) \le 1\times 10^{-3}$.
The measurement we report here is several orders of magnitude more sensitive than this indirect limit.
 We use our result to place constraints on $\Lambda_{NP}$/$g_{NP}^{2}$ of NP
processes that include LFV, where $g_{NP}$ is the coupling of the NP and ${\Lambda_{NP}}$ is the energy scale of the NP. 

Our sample of $\Upsilon(3S)$ meson data was collected with the \babar\  detector at the \pep2\  
asymmetric-energy \epem\ collider at the SLAC National Accelerator Laboratory. The detector was 
operated from 1999 to 2008 and collected data at the center-of-mass (CM) energies of the $\Upsilon(4S)$ (10.58~GeV), 
$\Upsilon(3S)$ (10.36~GeV), and $\Upsilon(2S)$ (10.02~GeV) resonances, as well as at energies in the vicinity of these resonances.
 In this paper  we describe a direct search for LFV decays in a sample of 122 million $\Upsilon$(3S) decays 
corresponding to an integrated luminosity of 27.96$\pm$0.17~fb$^{-1}$~\cite{ref:lumi} collected 
during 2008 (referred to as Run~7). Data collected at the $\Upsilon(4S)$ in 2007 (referred to as Run~6) 
with an integrated luminosity of 78.31$\pm$0.35~fb$^{-1}$~\cite{ref:lumi}, data taken 40~MeV below the $\Upsilon(4S)$
 resonance corresponding to
7.752$\pm$0.036~fb$^{-1}$~\cite{ref:lumi},
and data taken 40~MeV below the $\Upsilon(3S)$ resonance corresponding to 2.623$\pm$0.017~fb$^{-1}$~\cite{ref:lumi}
 constitute control samples.
These are used to evaluate non-resonant contributions to the background and to study systematic effects in a signal-free sample. 
We employ a blind analysis strategy~\cite{ref:blindanalysis} in which  0.93~fb$^{-1}$ of the $\Upsilon(3S)$ sample is used solely
 in the stage prior to unblinding, during which selection criteria are optimized and all systematic uncertainties evaluated.
The data sample reserved for the LFV search is based on (117.7$\pm$1.2)$\times$10$^{6}$ $\Upsilon(3S)$ decays,
 corresponding to 27.02$\pm$0.16~fb$^{-1}$, and excludes the 0.93~fb$^{-1}$ sample.

In the \babar\ detector, which is described in detail elsewhere~\cite{ref:aubert,ref:NIMUpdate}, the trajectories of
charged particles  are measured in a 5-layer silicon vertex tracker (SVT) 
surrounded by a 40-layer cylindrical drift chamber (DCH). This charged particle tracking system is inside
a 1.5~T solenoid with its field running approximately  parallel to the $e^+e^-$ beams and together they form
a magnetic spectrometer. In order to identify and measure the energies and directional information of photons and electrons,
 an electromagnetic calorimeter (EMC) composed of 
an array of 6580 thallium doped CsI crystals, located just inside the 
superconducting magnet, is used. Muons and neutral hadrons 
are identified by arrays of resistive plate chambers or limited steamer-tube 
detectors inserted into gaps in the steel of the Instrumented Flux Return (IFR) of the 
magnet. The $\Upsilon(4S)$ control sample data for this analysis are
restricted to Run~6 to ensure that the control (Run~6) and signal (Run~7) 
data sets have the same IFR detector configurations following an IFR upgrade 
program that was completed prior to the beginning of Run~6.

The signature for $\Upsilon(3S)$ $\rightarrow$ $e^{\pm}$$\mu^{\mp}$ events consists 
of exactly two oppositely charged primary particles, an electron and a muon, each with 
an energy close to half the total energy of the $e^+e^-$ collision in the CM frame, $E_B$.
There are two main sources of background:
(i) $e^+e^- \rightarrow \mu^{+}\mu^{-}(\gamma)$ events in which one of the muons is misidentified,
 decays in flight, or generates an electron in a material interaction;
and (ii) $e^+e^- \rightarrow e^{+}e^{-}(\gamma)$ events in which one of the electrons is misidentified.
Background from $e^+e^- \rightarrow  \tau^{+}\tau^{-} \rightarrow e^{\pm} \mu^{\mp}2\nu 2\bar{\nu}$
 is efficiently  removed with the kinematic requirements described below.
 Generic $\Upsilon(3S)$ decays to two charged particles where there is particle misidentification
are also a potential background.



The $\Upsilon(4S)$ Run~6 data, which is at a CM energy above the $\Upsilon(3S)$ mass,
 is used as a high statistics control sample to estimate the continuum, i.e., non-$\Upsilon(3S)$, background in Run~7 data.
Although collected at the $\Upsilon(4S)$ resonance, because the large width of the $\Upsilon(4S)\rightarrow B \bar{B}$ strong decays 
suppresses the branching fractions of lepton-pair $\Upsilon(4S)$ decays as well as any potential LFV decays,
Run~6 data provides a reliable  continuum control sample.
 The KK2F Monte Carlo (MC) generator~\cite{ref:ward} 
is used to simulate $\mu$-pair and $\tau$-pair events. 
The BHWIDE generator~\cite{ref:jadach} is used to simulate Bhabha events.
Both generators take into account initial- and final-state radiation.
They also are used to cross-check the Run~6 data-driven estimates of the continuum background.
The EvtGen generator~\cite{ref:lange} is used to simulate hadronic 
continuum events and generic $\Upsilon(3S)$ decays, as well as the
 signal $\Upsilon (3S)\rightarrow e^{\pm}\mu^{\mp}$ decays, in which the electron and muon have a $(1+\cos^2\theta)$ 
distribution, where $\theta$ is the CM polar angle relative to the $e^-$ beam.
 The simulated $\mu$-pair, $\tau$-pair and generic $\Upsilon(3S)$ samples 
correspond to approximately twice the number of events 
expected in the $\Upsilon(3S)$ data set, while the Bhabha sample corresponds to approximately half the number 
of events. The GEANT4~\cite{ref:agostine} suite of programs
 is used to simulate the response of the \babar\ detector.

Event selection proceeds in two stages. In the first stage, a dedicated \emu 
 filter is used to preselect events with only
 an electron candidate and a muon candidate in the detector. 
 In this filter all events,
in addition to passing either the drift chamber or electromagnetic calorimeter higher level triggers,
 are required to have exactly two tracks of opposite charge that are separated by more than 90$^{\circ}$ 
 in the CM.
 One of the tracks must pass a very loose electron selection ($E/p>0.8$) and the other a very loose muon requirement ($E/p<0.8$), 
where $E$ is the energy deposited in the EMC  associated with the track of momentum $p$.
 The preselection has an 80\% efficiency for signal events, including geometrical acceptance.
The first row in Table~\ref{tab:tb1} documents this preselection efficiency along with the
 numbers of background events expected from generic $\Upsilon(3S)$ decays, as predicted by EvtGen, and
 the amount of background  from  the continuum as determined by the Run~6 data control sample.
 It also includes the number of  events  preselected from the unblinded $\Upsilon$(3S) data sample.


\begin{table}[h]
\begin{center}
\caption {Impact of each component of the selection on the signal efficiency, number of background events, and
number of events in the data. The first row provides information on the pre-selection. The last row provides the information
after applying all selection criteria. Rows 2 to 7 provide information when all requirements are applied
except the criterion associated with the particular row. The luminosity-normalized expected number of events in the third and fourth columns 
are the background events from the generic $e^{+}e^{-}\rightarrow\Upsilon(3S)$ MC and
the data-driven continuum background events estimated from the $e^{+} e^{-}\rightarrow \Upsilon(4S)$
 sample, respectively.
The last column represents the number of events in the $\Upsilon(3S)$ data sample after unblinding.}
\vspace{5 mm}
\label{tab:tb1}
\begin{tabular}{|c|c|c|c|c|} \hline
\centering
 
\textbf{Selection}     & \textbf{Signal}     &\textbf{$\Upsilon(3S)$}    & \textbf{Continuum}            & \textbf{Events}    \\   
\textbf{Criterion}     & \textbf{Efficiency} ($\%$) &     \textbf{BG}     & \textbf{BG}        & \textbf{in Data}          \\ \hline \hline

Pre-Selec.             & ~~~80.20                  & ~75516                     & 725003                    & 945480                   \\ 
                       & $\pm$ 0.12            & $\pm$ 180                 & $\pm$ 500                 &                          \\ \hline \hline

Optimized              & ~~~50.74                  & ~5180                         & 320910                 & 358322                   \\                  
PID                    & $\pm$ 0.15            & $\pm$ 50                  & $\pm$ 330                 &                          \\ \hline 

2 tracks               & ~~~23.54                  & 0                            & ~~14.1                        & 18                       \\ 
in final               & $\pm$ 0.13            &                              & $\pm$ 2.2                   &                             \\ 
state                  &                         &                              &                              &                          \\ \hline

Lep. Mom.              & ~~~26.84                  & ~~87                        & 253                       & 302                      \\ 
                       & $\pm$ 0.12            & $\pm$ 6                   & $\pm$ 9                   &                          \\ \hline
 
Back-to-               & ~~~24.02                  & ~~~0.5                         & ~~36                        & 39                       \\ 
back                   & $\pm$ 0.13            & $\pm$ 0.5                   & $\pm$ 6                   &                          \\ \hline
  
EMC                    & ~~~24.95                  & 0                            & ~~13.5                        & 17                       \\ 
Accept.                & $\pm$ 0.13            &                              & $\pm$ 2.2                  &                          \\ \hline
    
Energy on              & ~~~24.52                  & 0                            & ~~16.9                        & 19                       \\ 
EMC                    & $\pm$ 0.13            &                              & $\pm$ 2.4                   &                          \\ \hline \hline
  
All Criteria           & ~~~23.42                  & 0                            & ~~12.2                         & 15                       \\ 
                       & $\pm$ 0.13            &                              & $\pm$ 2.1                    &                          \\ \hline
\end{tabular}
\end{center}
\end{table}

In the second stage of the analysis, we apply tighter and optimized particle identification (PID) and kinematic criteria.
Applying PID to select events with one muon and one electron is the most effective means of reducing the background
while maintaining an acceptable efficiency. 
All components of the \babar\ detector contribute to PID. 
Different PID selectors have been developed by \babar\
to distinguish each particle type in a set of multivariate analyses.
 These are described in more detail in Ref.~\cite{ref:NIMUpdate}.
Selectors for this analysis based on  error-correcting output codes~\cite{ref:KM}
 and decision trees are used to identify electrons, pions, and kaons, whereas
 bagged decision tree selectors~\cite{ref:narsky} are used to identify muons.
 The selectors can be deployed to provide different nested levels of particle efficiency and background, 
from ``Super Loose'' (most efficient, least pure) to ``Super Tight'' (least efficient, most pure),
 where candidates selected by tighter criteria of the selector are a subset of events 
that pass looser selection criteria.
We optimize the choice  of the electron and muon selectors
to maximize $\varepsilon_{e\mu}/\sqrt{1+N_\mathrm{BG}}$,
 where $\varepsilon_{e\mu}$ is the final efficiency as determined from 
signal MC, and $N_\mathrm{BG}$ is the number of expected background events 
as predicted by data control samples in Run~6 and generic $\Upsilon(3S)$ 
MC events.
In the optimized selection, electron candidates are required to pass the ``Super Tight'' electron selector and
muon candidates to pass the ``Tight'' muon selector. In addition, 
electron candidates are required to fail the ``Tight'' muon selector and muons are required to fail the ``Super Tight'' electron selector.
Each track is also required to fail the ``Loose'' pion selector as well as the ``Loose'' kaon selector.

\begin{figure}[!htb]
\begin{center}
\includegraphics[height=5.2cm,width=8.7cm]{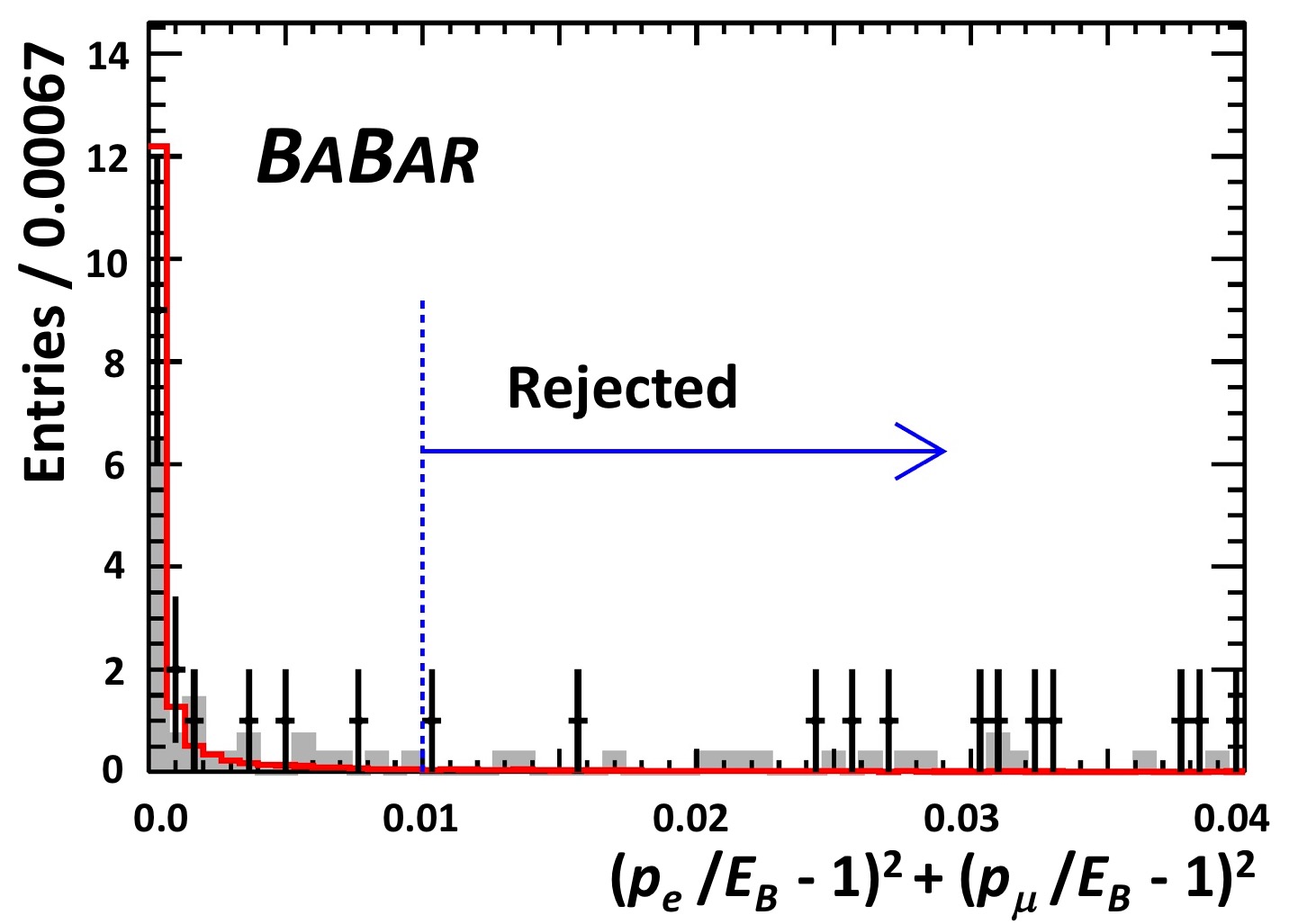}   
\caption{
Distribution of the squared distance from the point (1,1) in the plane of the scaled electron {\it vs} muon momenta
 for events satisfying all other selection criteria in the data (points with error bars),
 continuum background (gray histogram, from the Run~6 control sample  normalized to 27.02~fb$^{-1}$),
 and simulated signal (red histogram, arbitrarily normalized). 
 The dashed line indicates the position of the 0.01 requirement.
}
\label{fig:finalLepMomCir}
\end{center}
\end{figure}

Kinematic requirements are also applied to suppress
$e^+e^- \rightarrow  \tau^{+}\tau^{-} \rightarrow e^{\pm} \mu^{\mp}2\nu 2\bar{\nu}$ events,
radiative  Bhabha and $\mu$-pair events, the $e^+e^- \rightarrow e^+e^- e^+e^-$ 
and $e^+e^- \rightarrow e^+e^- \mu^+\mu^-$ two-photon processes, and beam-gas interactions. In the $p_e/E_B$ {\it vs} $p_{\mu}/E_B$ plane, where $p_e/E_B$~($p_{\mu}/E_B$) is the ratio of the 
electron (muon) momentum to the beam energy in the CM frame, the distribution of $e$-$\mu$ 
signal events peaks at (1,1). Events must lie within a circle about that peak: namely we require $(p_e/E_B-1)^2 + (p_{\mu}/E_B-1)^2 < 0.01$. 
Figure~\ref{fig:finalLepMomCir} shows the distribution of $(p_e/E_B-1)^2 + (p_{\mu}/E_B-1)^2$
after all other selection criteria have been applied,
for the signal, $\Upsilon(3S)$ data sample, and continuum backgrounds estimated from Run~6 data.
 The angle between the two lepton tracks in the CM is required to be more than 179$^{\circ}$. In order to reduce continuum background from $\mu$-pairs and to suppress Bhabha events in which an electron is misidentified as a muon because it passes through the space between crystals, the primary muon candidate is required to deposit at least 50~MeV in the EMC. We require that the lepton tracks fall within the angular 
acceptance (24$^{\circ}$ $<$ $\theta_{\mathrm{Lab}}$ $<$ 130$^{\circ}$) of the EMC, 
where $\theta_{\mathrm{Lab}}$ is the polar angle of lepton tracks in the lab frame.

\begin{figure}[!htb]
\begin{center}
\includegraphics[height=5.5cm,width=8.7cm]{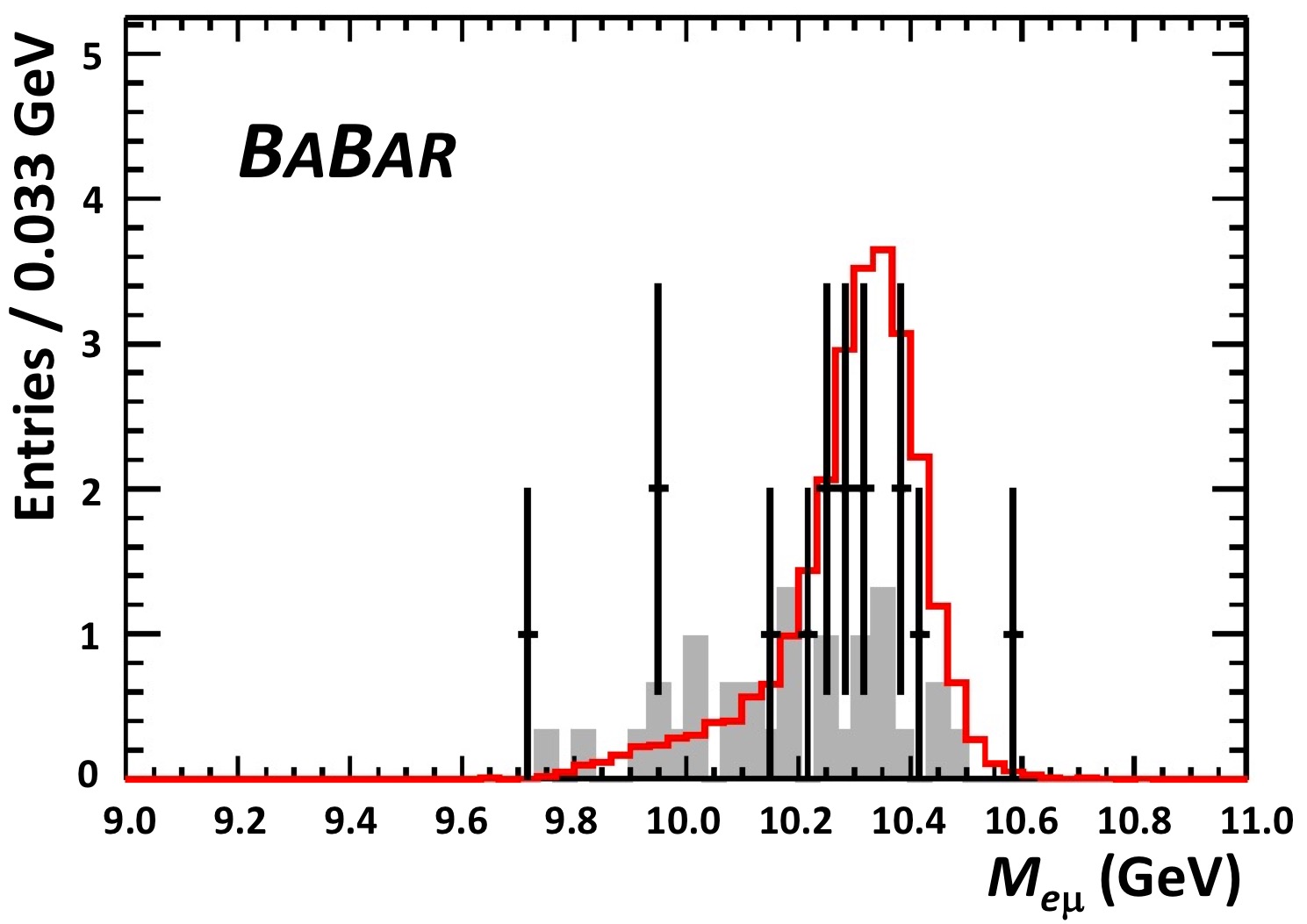}  
\caption{The distribution of the \emu 
 invariant mass of events surviving all selection criteria.
The data sample is presented as the histogram in black with error bars and the open red 
histogram represents the signal MC with arbitrary normalization.
 The grey histogram shows the estimate of the contiuum background
from the Run~6 control sample data with the rate scaled to the amounts expected at 10.36~GeV for a data sample of 27.02~fb$^{-1}$
and the mass scaled to 10.36/10.58.
}
\label{fig:final_emu_mass}
\end{center}
\end{figure}

The signal efficiency, as determined from signal MC, is (23.42$\pm$0.13\,(stat))$\%$. 
Figure~\ref{fig:final_emu_mass} shows the \emu  
invariant mass distribution of the 
data candidates and background events, as well as the potential signal, after all selection requirements have been applied.

No events from the generic $\Upsilon(3S)$ MC sample survive the selection. 
 We estimate an uncertainty of ${\pm}0.9$ events in this source of potentially misidentified generic $\Upsilon(3S)$ decays
 by loosening the PID selectors and use the uncertainty in the 
surviving number of events with this loosened selection as the uncertainty in this background.
The determination of the continuum backgrounds obtained
using the Run~6 data, as described above,
predicts a background of 12.2$\pm$2.1  events from continuum processes.
The MC samples of the continuum are only used as a cross check on 
backgrounds at various stages of the analysis.
 No events from Bhabha, $\tau$-pairs, 
$c\bar{c}$, $u\bar{u}+d\bar{d}+s\bar{s}$,
 or generic $\Upsilon(3S)$ MC pass the selection. The MC 
predicts that 0.16$\pm$0.05 continuum $\mu$-pair events survive the final selection. 
 An uncertainty of ${\pm}$2.3 events is assigned to the total background estimate,
calculated as the quadrature sum of the uncertainties in the $\Upsilon(3S)$ and
continuum background estimates.

Table \ref{tab:tb1} summarizes the signal efficiency, estimates of the numbers of background events,
and numbers of events in the $\Upsilon(3S)$ data sample at the various stages of the selection.

\begin{figure}[!htb]
\begin{center}
\includegraphics[height=5.5cm,width=8.65cm]{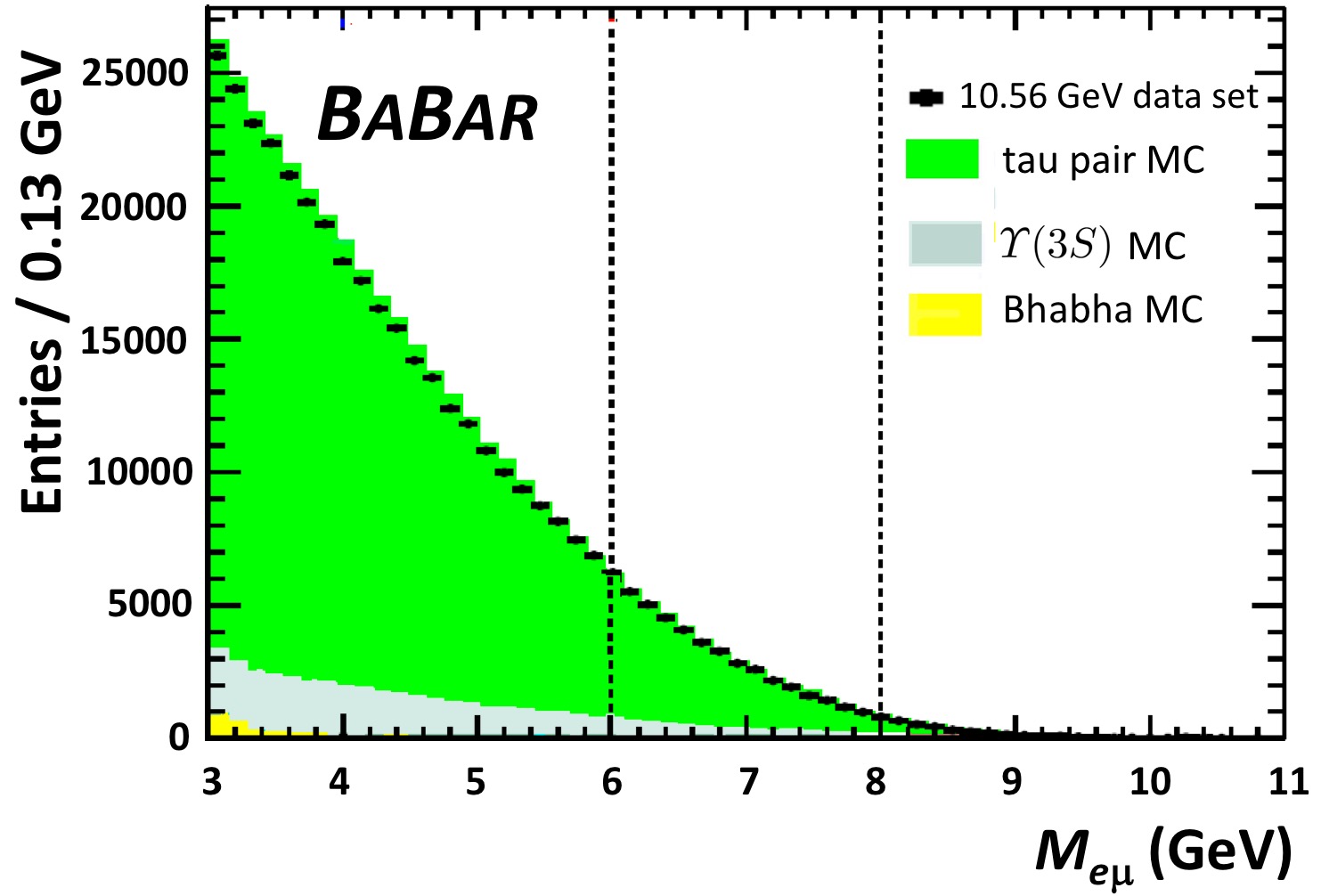} 
\caption{The distribution of the \emu 
invariant mass of events 
in a control sample  dominated by $\tau$-pair events obtained by reversing the  
two major kinematic requirements in the selection.
%
 The green, light grey, and yellow colored histograms represent the MC
predictions for  $\tau^+\tau^-$, generic $\Upsilon(3S)$ 
decays, and Bhabha events, respectively, while the histogram represented by the black line 
with error bars represents data from the $\Upsilon(3S)$ data sample.
The systematic uncertainty in the signal efficiency associated with all requirements
but those on the two kinematic requirements used to define this sample
is obtained by comparing the MC expectations with the data in the 
side-band region indicated by the dashed vertical lines.
}
\label{fig:sidebands}
\end{center}
\end{figure}

We assess the systematic uncertainties in the signal efficiency by determining the ratio of data to MC
yields for a control sample of $e^+e^-\rightarrow \tau^+\tau^- \rightarrow e^{\pm}\mu^{\mp} 2\nu 2\bar{\nu}$
 events in an \emu 
  mass sideband. For this study,
 we reverse the two major kinematic requirements, the $E_B$-normalized lepton momentum cut and 
the requirement on the angle between the two tracks, in order to obtain a large control sample of $\tau$-pair events.
This $\tau$ control sample study measures the systematic uncertainty associated with particle identification, tracking,
 kinematics, trigger selection criteria, and all other effects except those associated with the two major kinematic requirements
 used to select the control sample.
 Figure~\ref{fig:sidebands} shows the distribution of $M_{e\mu}$ for the data and MC in the $\tau$ control sample.
 We evaluate the associated correction to the signal efficiency by measuring the ratio
$(N_{\mathrm{Data}} - N^{\mathrm{MC}}_{non-\tau^+\tau^-})/N^{\mathrm{MC}}_{\tau^+\tau^-}$
 in the sideband region 6~GeV$<$$M_{e\mu}$$<$8~GeV, which is just below our signal region,
 where $N_{\mathrm{Data}}$ is the number of events in the data, 
 $N^{\mathrm{MC}}_{\tau^+\tau^-}$ is the number of  $\tau$-pair events predicted in MC, and
$N^{\mathrm{MC}}_{non-\tau^+\tau^-}$ the number of MC-predicted events that are not $\tau$-pairs.
 We obtain a value of 1.007$\pm$0.010(stat) for this ratio. We take the quadratic sum of the statistical uncertainty
in this ratio and difference from unity as this part of the relative systematic uncertainty, 1.2\%, in the signal efficiency. 
 We evaluate the systematic uncertainties associated with the
 two major kinematic requirements that are reversed for the $\tau$ control sample selection
 by using them to select a $\mu$-pair control sample having very similar
kinematic properties as the signal. We conservatively vary the values of 
the two selection criteria from the default values and assign the differences in the
 selection efficiencies between MC and data for the $\mu$-pair control samples as the relative efficiency
  uncertainties associated with these requirements.
 The $E_B$-normalized momentum requirement is varied by ${\pm}$0.0015 and the back-to-back angle requirement by ${\pm}$0.1$^{\circ}$. 
The number of signal events remains unchanged under these variations.
Table~\ref{tab:tb2} summarizes the systematic uncertainties.
The signal efficiency is (23.4$\pm$0.8)$\%$, where the quoted uncertainty is determined by
 summing in quadrature the individual contributions.


\begin{table}[!htb]
\caption{Summary of systematic uncertainties. 
The values of the efficiency, background, and number of $\Upsilon(3S)$ decays
are presented in the first column and their uncertainties in the second column.
The different contributions to the efficiency systematic uncertainties are also
 presented.}
\begin{center}
\vspace{5 mm}
\label{tab:tb2}
\begin{tabular}{|l|l|l|} \hline
\centering
\textbf{Component Value}            & \multicolumn{2}{c|}{\textbf{Uncertainties by Source}}       \\ \hline
                                    & Lep. Mom. cut:          & 0.0068 (2.9 $\%$)    \\
                                    & Back-to-back cut:       & 0.0026 (1.1 $\%$)     \\
                                    & All other cuts:         & 0.0028 (1.2 $\%$)     \\
Signal                              & MC statistics:          & 0.0003 (0.13 $\%$)    \\ \cline{2-3}
Efficiency: 0.2342                  & \multicolumn{2}{c|}{$\pm 0.0078$ (3.3 $\%$)} \\ \hline \hline

${N_{\Upsilon}}$: 117.7 $\times$ 10$^{6}$ & \multicolumn{2}{c|}{${\pm}1.2\times 10^{6}$  (1.0 $\%$)}  \\ \hline \hline
BG: 12.2                               &  \multicolumn{2}{c|}{ ${\pm}2.3$ (19 $\%$)}                       \\ \hline 

\end{tabular}
\end{center}
\end{table}

After unblinding the data, we find $N_{\mathrm{cand}}$=15 candidate events and have an
expected background of 12.2${\pm}$2.3 events 
from a sample of (117.7$\pm$1.2)$\times 10^6$ $\Upsilon(3S)$ mesons. 
Calculating the branching fraction from
$(N_\mathrm{cand} - N_\mathrm{BG})/ (\varepsilon_{sig} N_{\Upsilon(3S)})$ gives:
\begin{equation}
\mathcal{B}(\Upsilon (3S)\rightarrow e^{\pm}\mu^{\mp}) =
(1.0 \pm 1.4 \, \mathrm {(stat)} \pm 0.8 \, \mathrm {(syst)})\times 10^{-7}
\end{equation}
where the statistical uncertainty is that from $N_\mathrm{cand}$ and all other uncertainties
are included in the systematic uncertainty.
As  this result is consistent with no signal, we set an upper limit
at 90$\%$ confidence level (CL)  on the branching fraction by
using the CLs method~\cite{ref:CLs}, a modified frequentist method
that accomodates potential large downward fluctuations 
in backgrounds:
\begin{equation}
\mathcal{B}(\Upsilon (3S)\rightarrow e^{\pm}\mu^{\mp})<3.6\times 10^{-7}~@~90\%~\mathrm{CL}.
\label{ULResult}
\end{equation}
The CLs expected 90\%~CL upper limit, given the number of background events and assuming no signal, is 2.8$\times 10^{-7}$.

This result is the first reported experimental upper limit  on 
$\Upsilon (3S)\rightarrow e^{\pm}\mu^{\mp}$ and from any $b{\bar b}$ bound state.
It can be interpreted as  a limit on NP using the relationship 
$\left( g_{NP}^{2}/\Lambda_{NP}\right)^2 / \left( 4 \pi\alpha_{3S} Q_b/M_{\Upsilon(3S)}\right)^{2} = 
\mathcal{B}(\Upsilon(3S)\rightarrow e\mu)/\mathcal{B}(\Upsilon(3S)\rightarrow \mu \mu)$,
ignoring small kinematic factors,
and where $Q_b=-1/3$ is the $b$-quark charge and $\alpha_{3S}$ is the fine structure constant at the $M_{\Upsilon(3S)}$ energy scale.
Using the world average $\mathcal{B}(\Upsilon (3S)\rightarrow \mu^+\mu^-)=2.18\pm 0.21$~\cite{Zyla:2020zbs}
 gives a 90$\%$~CL upper limit
 of $\Lambda_{NP}$/$g_{NP}^{2}$$>$80~TeV.

We are grateful for the excellent luminosity and machine conditions
provided by our \pep2\ colleagues, 
and for the substantial dedicated effort from
the computing organizations that support \babar.
The collaborating institutions wish to thank 
SLAC for its support and kind hospitality. 
This work is supported by
DOE
and NSF (USA),
NSERC (Canada),
CEA and
CNRS-IN2P3
(France),
BMBF and DFG
(Germany),
INFN (Italy),
FOM (The Netherlands),
NFR (Norway),
MES (Russia),
MINECO (Spain),
STFC (United Kingdom),
BSF (USA-Israel). 
Individuals have received support from the
Marie Curie EIF (European Union)
and the A.~P.~Sloan Foundation (USA).



\end{document}